# *Numerical simulation of internal incompressible flows with enhanced variants of dissipative inlet/outlet conditions. Part 1: Mathematical formulations and solution methods.*

*Jacek Szumbarski*

*Institute of Aeronautics and Applied Mechanics, Warsaw University of Technology*

*24 Nowowiejska St., 00-655 Warszawa, POLAND*

*e-mail: jasz@meil.pw.edu.pl*

*Abstract*

*The paper presents numerical methods for unsteady flows of a viscous incompressible fluid in internal domains with many inlet/outlet sections. The novel variants of dissipative boundary conditions augmented by the inertia terms are used at the inlets/outlets of the flow domain. Volumetric flow rates or inlet/outlet average pressure are imposed as additional constrains imposed on a fluid motion. The variational formulations of the Stokes problem with such conditions and constrains are presented and the solution methods are proposed. These methods are based on superposition of appropriately defined auxiliary Stokes problems. Extension of the proposed methodology to the Navier-Stokes flows, based on the Operator-Integration-Factor-Splitting (OIFS) technique, is also described. Next, a nonlinear extension of the inertial-dissipative conditions is formulated and incorporated in the solution framework for the Navier-Stokes flows. Finally, an alternative approach to flow problems with such conditions, based on optimal control theory and adjoint evaluation of the goal functional gradient, is presented. A few remarks concerning numerical implementation issues are formulated.*

## 1. Introduction

Numerical modeling of unsteady incompressible flows in complex internal domain has been continuously a challenge for both theoreticians as well as practitioners in the field of Computational Fluid Dynamics (CFD). One of the main source of difficulties is formulation of appropriate boundary conditions at artificial boundaries of a computational domain. For internal flows these boundaries have usually a form of inlet and outlet sections which appear due to trimming of a large flow domain to a smaller computational one. Such domain truncation, common especially in numerical simulations of flows in cardiovascular and respiratory systems, is necessary to achieve high modeling efficiency, both in terms of required memory resources and computational time.

When a desirable truncation of a flow domain is radical, it may lead to substantial difficulties in formulation of physically plausible, mathematically consistent and numerically stable inlet/outlet conditions at artificial boundaries. Extensive research has been carried out in this respect for at least last two decades. The problem of efficient outflow conditions for the wake flow behind an obstacle, imposed on a downstream boundary of the flow domain, was addressed in [1]. In the context of internal flows, the do-nothing boundary conditions have been originally proposed and analyzed in [2]. The essence of these conditions was to enforce a desired time variation of the integral rather than local quantities, like flow rate or surface-averaged pressure. In order to deal with the flow problems with flux rate constrains, the method based of superposition of special solutions called "the flux carriers" was proposed. A few years later, the concept of the do-nothing conditions (called also the defective conditions) was further developed by Formaggia at al [3], who casted the flow-rate conditions as the constrains realizable by a priori unknown inlet/outlet averaged pressure values playing the role of the Lagrange multipliers. Such approach bypasses necessity of finding the flux carriers. Moreover, this alternative approach is based only on standard functional spaces and as such it is easier to implement within the framework of finite or spectral element methods. In the work [4], Quarteroni et al. demonstrated how do-nothing inlet/outlet conditions can be incorporated within the multiscale modeling framework applied to the cardiovascular system. The do-nothing approach has been also exploited by Szumbarski et al. [5,6] to simulate blood flows in the simple model of the Blalock-Taussig shunt. In this work, the general method for unsteady flows with mixed flow-rate/average-pressure inlet/outlet conditions was formulated and tested. An alternative formulation of the defective inlet/outlet conditions, derived from the Navier-Stokes equation with vorticity-based viscous term, was proposed by Veneziani and Vergara [7]. Further, the optimal control based approach to internal flow problems with flow-rate inlet/outlet conditions was proposed by Formaggia et al. [8]. The essence of this formulation is to treat a priori unknown inlet/outlet average pressures as control variables and find their time variations such that discrepancies between requested and actual flow rates are minimized. The proposed minimization is gradient-based and uses the adjoint technique to evaluate the gradient of the minimized functional.

More recent developments and applications of the defective boundary conditions include their extension to fluid-structure interaction problems [9] and non-Newtonian fluids [10]. Consistent inclusion of such conditions into the pressure-correction and velocity-correction schemes has been presented in [11,12]. In [13], open boundary conditions have been built into, so called, unconstrained Navier-Stokes formulation, where appropriately postulated boundary-value problem for Pressure Poisson Equation (PPE) is solved instead of explicit use of the continuity equation. In this approach, the open boundary conditions at the inlets/outlets are transformed into Dirichlet boundary conditions for the pressure field. The latest papers, like [14] and [15], deal also with nonlinearly enhanced formulations which are design to eliminate stability problems of the standard do-nothing conditions.

For the sake of convenient reading, let us briefly recall the concept of the do-nothing inlet/outlet conditions. When posed in strong form, the do-nothing conditions are in fact a certain variant of more general open (or pseudo-traction) conditions. The latter can be written as follows

$$-p\boldsymbol{n} + \nu \frac{\partial}{\partial \boldsymbol{n}} \boldsymbol{u} = \boldsymbol{g} \qquad (1.1)$$

where $p$, $\boldsymbol{u}$ and $\nu$ denote, respectively, pressure (divided by density), velocity and kinematic viscosity of a fluid. The symbol $\boldsymbol{n}$ denotes the external normal vector to the boundary. The normal derivative of the velocity field is obtained by applying the tensor $\nabla \mathbf{u}$ to the normal vector $\mathbf{n}$ and, in Cartesian coordinates, can be written as follows

$$\frac{\partial}{\partial \boldsymbol{n}} \boldsymbol{u} = (\nabla \boldsymbol{u})\boldsymbol{n} = \frac{\partial u_i}{\partial x_j} n_j \, \boldsymbol{e}_i \qquad (1.2)$$

The right-hand-side vector $\boldsymbol{g}$ in (1.1) is defined along a part of a boundary where the open condition is imposed and, in general, can depend on time. For the do-nothing variant, this vector is defined as $\boldsymbol{g} = -P(t)\boldsymbol{n}$, where the function $P$ is either given or it should be found to obtain assumed flow rate variation through the boundary. In the latter case, the functions $P$ plays the role of the Lagrange multiplier which should be found together with the whole flow field.

Note that the do-nothing conditions can be split into normal and tangent components. Indeed, (1.1) with $\mathbf{g} = -P(t)\mathbf{n}$ can be equivalently written as

$$\begin{cases} -p + \nu \mathbf{n} \cdot \frac{\partial}{\partial \mathbf{n}} \mathbf{u} = -P(t) \\ \mathbf{n} \times \frac{\partial}{\partial \mathbf{n}} \mathbf{u} = \mathbf{0} \end{cases} \quad (1.3)$$

Assume that a part of a boundary where (1.3) is imposed is flat, i.e., the normal vector $\mathbf{n}$ is the same at each boundary point. Then, it can be shown that the second conditions in the pair (1.3) implies that the normal derivative of the velocity components tangent to the boundary is equal zero. In other words, in such circumstances the do-nothing condition implies homogeneous Neumann condition imposed on the tangent component of the velocity field. The first (scalar) condition in the pair (1.3) can be re-written as follows

$$-p + \mathbf{n} \cdot (\nu \mathbf{D} \mathbf{n}) = -P(t) \quad (1.4)$$

where $\mathbf{D} = \frac{1}{2}[\nabla \mathbf{u} + (\nabla \mathbf{u})^T]$ is the deformation rate tensor. Indeed, the (1.4) follows form the fact that $\mathbf{n} \cdot (\mathbf{R}\mathbf{n}) = 0$, where $\mathbf{R} = \frac{1}{2}[\nabla \mathbf{u} - (\nabla \mathbf{u})^T]$ is the antisymmetric rotation tensor.

Note that the condition (1.4) has no direct physical sense. Although looking similar, the expression in the left-hand side of (1.4) does not describes the normal stress distribution - the factor of 2 is missing at the viscous term in the left-hand side of (1.4). For this reason, the open boundary conditions are sometimes referred to as the pseudo-traction conditions. Note also that the conditions (1.3) provides also pressure normalization, i.e. it sets the absolute level of pressure in the computational domain. Hence, it is often natural to assume that at a selected inlet/outlet the function $P(t) \equiv 0$.

Finally, it can be shown that the condition (1.4) implies that the function $P$ is equal to the section-averaged instantaneous pressure. Indeed, if the inlet/section is flat and surrounded by a solid wall then the boundary (line or surface) integral of the viscous term in (1.4) is equal zero, see for instance [2] or [5]. This interpretation makes the do-nothing condition particularly attractive for the modeling of internal flow in biological networks, as it facilitates merging of local full 3D simulations with global section-averaged-based (1D) models. The discussion of such hybrid approach can be found in [4] and [16] and references cited therein.

Still, applications exist where the standard form of the do-nothing conditions (1.3) is not sufficient for convenient modeling. Specifically, realistic modeling of flows in the respiratory system requires further generalization of (1.3) which would by-pass difficulties with estimation of pressure variations at the outlets from airway paths. Moreover, it is desirable to include an additional flow resistance that appears "in the bulk" when the air enters the terminal small-scaled part of the respiratory system outside the resolved flow domain. To meet these needs, the concept of open dissipative boundary conditions was put forward and elaborated in details in [17,18]. In the recent monographic publication [19], Maury shows how 3D simulations with such conditions can be combined with reduced lower-dimensional models to give a full quantitative description of air flow in the respiratory system.

In the current work, we take over the original idea of the dissipative open conditions and proposed generalization which accounts for inertial effects. Including these effects in the inlet/outlet conditions seems to be particularly appropriate for strongly unsteady flow conditions. Yet another generalization is to add nonlinear dissipation term. In terms of physics, this step is equivalent to replacement of the Darcy-like dissipation model by more general Darcy-Forchheimer model. In terms of computations, such generalization poses a challenge as it introduces one more source of nonlinearity in the flow model. In the paper, we show how this nonlinearity can be conveniently approached within the Operator-Integration-Factor-Splitting (OIFS) technique [20,21]. Eventually, we show how generalized inertial-dissipative inlet/outlet conditions can be used to solve a problem of an unsteady internal flow in a domain with arbitrary number of inlet/outlet sections and mixed flow-rate/average-pressure boundary conditions. We also point out to certain implementation issues related to inertial terms in the inlet/outlet conditions. Finally, we present an alternative flow-control based formulation of the flow problem with nonlinear dissipative inlet/outlet conditions. The proposed methodology follows that proposed by Formaggia at al. [8], i.e., it applies the adjoint-based evaluation of the gradient of an appropriately formulated goal functional.

The content of the paper is following. In Section 2, we formulate the initial-boundary value problem for the Stokes flow in the domain with arbitrary number of inlet/outlet sections. Two variants of the open dissipative conditions supplemented with terms modeling fluid inertia are presented. Additional constrains are imposed, namely, it is assumed that the volumetric flow rate at a chosen subset of the inlets/outlets must follow prescribed time variation. The physical meaning of the postulated inlet/outlet conditions is also addressed. In particular, it is shown that proposed variants of the inlet/outlet conditions are equivalent in this sense that they imply the same relation between the section-averaged pressure and the flowrate. In Section 3, the numerical method for the unsteady Stokes flow with augmented dissipative inlet/outlet conditions and flowrate constrains defined in Section 2 is described. In this method, the instantaneous velocity and pressure fields are constructed as linear combinations of solutions to appropriately defined auxiliary Stokes problems. In Section 4, the numerical methodology proposed for the Stokes flows in Sections 3 is extended for unsteady Navier-Stokes flows with flow rate constrains. This generalization is quite straightforward and it is based on the Operator-Integration-Factor-Splitting (OIFS) approach to nonlinear convective term. In Section 5, the inertial-dissipative boundary conditions are further generalized to a unsteady nonlinear Darcy-Forchheimer (DF) form which is meant to mimic inflow from or outflow into a porous structure. In the solution procedure, additional nonlinear terms are incorporated in the OIFS sub-step. An alternative formulation of the flow problem with DF conditions, based on the optimal control theory, is described in details in Section 6. Finally, short summary and outlook are included in Section 7.

## 2. Unsteady Stokes flows in internal domains with augmented inertial-dissipative inlet/outlet conditions

In this section, we consider an unsteady Stokes flow problem in the domain with many inlet/outlet sections and formulate the solution method based on the instantaneous superposition of solution to auxiliary problems. We used generalized inertial-dissipative conditions imposed at inlets and outlets.

### 2.1. Mathematical formulation of the basic flow problem

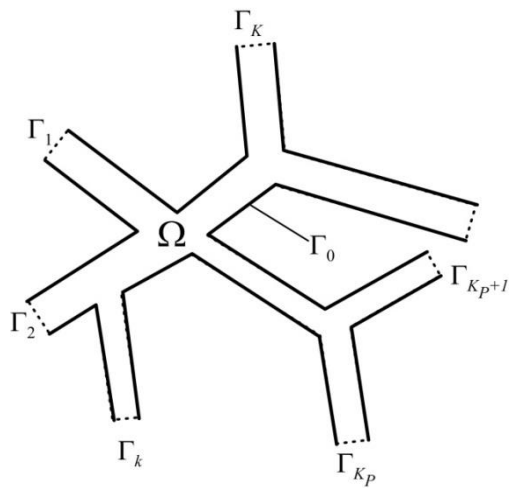

*Figure 1. An internal flow domain with multiply inlets/outlet.*

Consider the flow domain $\Omega$ in the form of the duct system depicted schematically in Figure 1. The boundary $\partial\Omega$ of this domain consists of the rigid impermeable wall $\Gamma_0$ and the inlet/outlet sections $\Gamma_k, k=1,..,K$. Consider next an unsteady Stokes problem posed in the domain $\Omega$

$$\begin{cases} \partial_t \mathbf{u} - \nu\Delta\mathbf{u} + \nabla p = \mathbf{f} \\ \nabla\cdot\mathbf{u} = 0 \end{cases} \qquad (2.1)$$

where the symbols $\mathbf{u}$, $p$, $\nu$ and $\mathbf{g}$ denote, respectively, velocity, pressure-to-density ratio, kinematic viscosity and the external mass-specific force field. The boundary condition at the solid wall $\Gamma_0$ is $\mathbf{u}\big|_{\Gamma_0} = \mathbf{0}$. The inlet/outlet conditions at the sections $\Gamma_k$, $k=1,..,K$, are imposed in any of the following variants:

(A) $\quad p\mathbf{n} - \nu\frac{\partial}{\partial \mathbf{n}}\mathbf{u} - \mathbf{n}(\lambda_k + \gamma_k \frac{d}{dt})\Phi_k(\mathbf{u}) = S_k\mathbf{n}\ at\ \Gamma_k,\ k=1,..,K$  (2.2a)

(B) $\quad p\mathbf{n} - \nu\frac{\partial}{\partial \mathbf{n}}\mathbf{u} - \mathbf{n}(\hat{\lambda}_k + \hat{\gamma}_k \partial_t)(\mathbf{u}\cdot\mathbf{n}) = S_k\mathbf{n}\ at\ \Gamma_k,\ k=1,..,K$  (2.2b)

where the symbol $\Phi_k$ denotes the flowrate operator, i.e., $\Phi_k(\mathbf{u}) = \int_{\Gamma_k} \mathbf{u}\cdot\mathbf{n}\,dA$ and the coefficients $\{\lambda_k, \gamma_k\}$ in (2.2a) or $\{\hat{\lambda}_k, \hat{\gamma}_k\}$ in (2.2b) are given for all $k=1,..,K$. The initial condition for the velocity field is $\mathbf{u}(t=0) = \mathbf{u}_0$.

Further, we assume that also given are the functions $\{S_1(t),..,S_{K_P}(t)\}$, while the functions $\{S_{K_P+1}(t),..,S_K(t)\}$ should be found so that the volumetric flow rates through the inlets/outlets $\{\Gamma_k, k=K_P+1,..,K\}$ follow assumed time variations, namely

$$\Phi_k(\mathbf{u}) = Q_k(t)\ ,\ k=K_P+1,..,K \qquad (2.3)$$

Note that the boundary conditions (2.2a) are the dissipative inlet/outlet conditions augmented with the inertia term when $\gamma_k \neq 0$. The variant B of the inlet/outlet conditions uses the inlet/outlet distribution of normal velocity component rather than volumetric flow rate. We will see later that these conditions have "in the mean" the same effect on the flow as the conditions in the variant A, but the corresponding variational formulation of the Stokes problem is different.

## 2.2. Physical meaning of the boundary conditions

Assume that the surface of an inlet/outlet $\Gamma_k$ to the flow domain $\Omega$ is flat, i.e., the normal vector is the same at each point of $\Gamma_k$. Then, it can be shown that

$$\int_{\Gamma_k} \mathbf{n} \cdot \frac{\partial}{\partial \mathbf{n}} \mathbf{u} \, dA = 0 \tag{2.4}$$

Indeed, choosing conveniently such reference frame that $\mathbf{e}_1 = \mathbf{n}|_{\Gamma_k}$ one obtains

$$\int_{\Gamma_k} \mathbf{n} \cdot \frac{\partial}{\partial \mathbf{n}} \mathbf{u} \, dS = \int_{\Gamma_k} \frac{\partial u_1}{\partial x_1} dA = -\int_{\Gamma_k} \left( \frac{\partial u_2}{\partial x_2} + \frac{\partial u_3}{\partial x_3} \right) dA = -\oint_{\partial \Gamma_k} (u_2 \eta_2 + u_3 \eta_3) dl = 0 \tag{2.5}$$

In (2.5), we have used the continuity condition and the Gauss-Ostrogradsky theorem in the plane $0x_2 x_3$. The symbol $\partial \Gamma_k$ denotes the contour line of the inlet/outlet section $\Gamma_k$ and $\boldsymbol{\eta} = [0, \eta_2, \eta_3]$ is the unit vector normal to this line. Clearly, the contour integral in (2.5) vanishes since the contour $\partial \Gamma_k$ lies on the solid side wall $\Gamma_0$.

Note that the inlet/outlet conditions in the form (2.2a) imply that at each $\Gamma_k$ one has

$$\begin{cases} p - \nu \mathbf{n} \cdot \frac{\partial}{\partial \mathbf{n}} \mathbf{u} - (\lambda_k + \gamma_k \frac{d}{dt}) \Phi_k(\mathbf{u}) = S_k \\ \mathbf{n} \times \frac{\partial}{\partial \mathbf{n}} \mathbf{u} = \mathbf{0} \end{cases} \tag{2.6}$$

Integration over $\Gamma_k$ of the first condition in (2.6) followed by division by the surface area $|\Gamma_k|$ yields

$$\bar{p}_k - S_k = (\lambda_k + \gamma_k \frac{d}{dt}) \Phi_k(\mathbf{u}) \tag{2.7}$$

Note that the viscous term has vanished due to (2.4).

The quantity $\bar{p}_k = \frac{1}{|\Gamma_k|}\int_{\Gamma_k} p\, dA$ is the average pressure at the inlet/outlet section $\Gamma_k$. Hence, the quantities $S_k$ can be interpreted as, generally time-dependent, pressure in the "far-field container" which supplies or consumes the flow rate through the section $\Gamma_k$. The pressure difference between the far-field and the inlet/outlet is expressed as the sum of two components: the "static" or "resistant" component which is proportional to the instantaneous flowrate $\Phi_k(\mathbf{u})$ and the "dynamic" or "inertia" component which is proportional to the rate of change (time derivative) of the flowrate. Addition of the latter term seems to be particularly appropriate for flows with pronounced unsteadiness, e.g. strongly pulsatile or oscillatory flows.

If the variant B of the inlet/outlet conditions is applied, the formulae analogical to (2.6) are

$$\begin{cases} p - \nu \mathbf{n}\cdot\frac{\partial}{\partial \mathbf{n}}\mathbf{u} - (\hat{\lambda}_k + \hat{\gamma}_k \frac{d}{dt})(\mathbf{u}\cdot\mathbf{n}) = S_k \\ \mathbf{n}\times\frac{\partial}{\partial \mathbf{n}}\mathbf{u} = \mathbf{0} \end{cases} \qquad (2.8)$$

Integration over $\Gamma_k$ of the first condition in (2.8) followed by division by the surface area $|\Gamma_k|$ yields

$$\bar{p}_k - S_k = |\Gamma_k|^{-1}(\hat{\lambda}_k + \hat{\gamma}_k \frac{d}{dt})\Phi_k(\mathbf{u}) \qquad (2.9)$$

Equivalence between (2.7) and (2.9) is evident with $\lambda_k = |\Gamma_k|^{-1}\hat{\lambda}_k$ and $\gamma_k = |\Gamma_k|^{-1}\hat{\gamma}_k$.

## 3. The solution method

In this section, we describe in details the method of solution of the unsteady Stokes problem (2.1) with the inlet/outlet conditions (2.2a) or (2.2b), and the flowrate constrains (2.3). First, we will pose problem in a weak form.

To this aim, we introduce the velocity functional space $V = \{\mathbf{v} : \mathbf{v} \in [H^1(\Omega)]^d, \mathbf{v}|_{\Gamma_0} = \mathbf{0}\}$. We also defined the bi-linear boundary functionals:

$$b_k(\mathbf{u}, \mathbf{v}) = \begin{cases} \left(\lambda_k + \gamma_k \frac{d}{dt}\right) \Phi_k(\mathbf{u}) \Phi_k(\mathbf{v}) & \text{in the variant A} \\ \left(\hat{\lambda}_k + \hat{\gamma}_k \frac{d}{dt}\right) \int_{\Gamma_k} (\mathbf{u} \cdot \mathbf{n})(\mathbf{v} \cdot \mathbf{n}) \, dA & \text{in the variant B} \end{cases}, \quad k = 1,..,K \quad (3.1)$$

Then the Stokes problem (2.1)-(2.3) can be formulated in weak sense as follows:

*Find:* $\mathbf{u} \in V$ *and* $p \in L^2(\Omega)$ *and* $\{S_{K_P+1},..,S_K\}$ *such that the equalities*

$$\begin{cases} \int_\Omega \partial_t \mathbf{u} \cdot \mathbf{v} \, d\Omega + \nu \int_\Omega \nabla \mathbf{u} : \nabla \mathbf{v} \, d\Omega - \int_\Omega p \nabla \cdot \mathbf{v} \, d\Omega + \sum_{k=1}^K \left[b_k(\mathbf{u}, \mathbf{v}) + S_k \Phi_k(\mathbf{v})\right] = \\ = \int_\Omega \mathbf{f} \cdot \mathbf{v} \, d\Omega \\ \int_\Omega q \nabla \cdot \mathbf{u} \, d\Omega = 0 \end{cases} \quad (3.2)$$

*hold true for each* $\mathbf{v} \in V$, $q \in L^2(\Omega)$, *and* $\Phi_k(\mathbf{u}) = Q_k(t)$ *for* $k = K_P + 1,..,K$. *In the above,* $\Phi_k(\mathbf{v}) \equiv \int_{\Gamma_k} \mathbf{v} \cdot \mathbf{n} \, dA$, $k = 1,..,K$.

A time-discretized form of the variational equalities (3.2) will be obtained as a next step. To this aim, we approximate local time derivative of the velocity field **u** by the backward-differentiation formula of the *J*-th order

$$\partial_t \mathbf{u}|_{t=t_{n+1}} \approx \frac{\alpha_0}{\Delta t} \mathbf{u}^{n+1} - \frac{1}{\Delta t} \sum_{j=1}^J \alpha_j \mathbf{u}^{n+1-j}$$

where the coefficients $\alpha_0,..,\alpha_J$ for different orders *J* are provided in the Table 1.

|       | $\alpha_0$ | $\alpha_1$ | $\alpha_2$ | $\alpha_3$ | $\alpha_4$ |
|-------|------------|------------|------------|------------|------------|
| *J=1* | 1          | 1          | -          | -          | -          |
| *J=2* | 3/2        | 2          | -1/2       | -          | -          |
| *J=3* | 11/6       | 3          | -3/2       | 1/3        | -          |
| *J=4* | 25/12      | 4          | -3         | 4/3        | -1/4       |

*Table 1. Coefficient of the BDF schemes of different orders.*

Then, the semi-discretized form of the variational problem (3.2) is

$$\begin{cases} \dfrac{\alpha_0}{\Delta t}\int_\Omega \mathbf{u}^{n+1}\cdot \mathbf{v}\,d\Omega + \nu\int_\Omega \nabla\mathbf{u}^{n+1}:\nabla\mathbf{v}\,d\Omega - \int_\Omega p^{n+1}\nabla\cdot\mathbf{v}\,d\Omega + \\ +\sum\limits_{k=1}^{K}\left[B_k(\mathbf{u}^{n+1}) + S_k^{n+1}\Phi_k(\mathbf{v})\right] = \mathbf{G}^{n+1} \quad , \quad \forall \mathbf{v}\in V \\ \int_\Omega q\nabla\cdot\mathbf{u}^{n+1}\,d\Omega = 0 \quad , \quad \forall q\in L^2(\Omega) \end{cases} \quad (3.3)$$

where:

$$B_k(\mathbf{u}^{n+1},\mathbf{v}) = \begin{cases} \left(\lambda_k + \gamma_k\dfrac{\alpha_0}{\Delta t}\right)\Phi_k(\mathbf{u}^{n+1})\Phi_k(\mathbf{v}) & : variant\ A \\ \left(\hat{\lambda}_k + \hat{\gamma}_k\dfrac{\alpha_0}{\Delta t}\right)\int_{\Gamma_k}(\mathbf{u}^{n+1}\cdot\mathbf{n})(\mathbf{v}\cdot\mathbf{n})\,dA & : variant\ B \end{cases} \quad (3.4)$$

$$\mathbf{G}^{n+1} = \begin{cases} \int_\Omega \mathbf{f}^{n+1}\cdot\mathbf{v}\,d\Omega + \dfrac{1}{\Delta t}\sum\limits_{j=1}^{J}\alpha_j\int_\Omega \mathbf{u}^{n+1-j}\cdot\mathbf{v}\,d\Omega + \\ \quad +\dfrac{1}{\Delta t}\sum\limits_{j=1}^{J}\alpha_j\sum\limits_{k=1}^{K}\gamma_k\Phi_k(\mathbf{u}^{n+1-j})\Phi_k(\mathbf{v}) & : variant\ A \\ \int_\Omega \mathbf{f}^{n+1}\cdot\mathbf{v}\,d\Omega + \dfrac{1}{\Delta t}\sum\limits_{j=1}^{J}\alpha_j\int_\Omega \mathbf{u}^{n+1-j}\cdot\mathbf{v}\,d\Omega + \\ \quad +\dfrac{1}{\Delta t}\sum\limits_{j=1}^{J}\alpha_j\sum\limits_{k=1}^{K}\hat{\gamma}_k\int_{\Gamma_k}(\mathbf{u}^{n+1-j}\cdot\mathbf{n})(\mathbf{v}\cdot\mathbf{n})\,dA & : variant\ B \end{cases} \quad (3.5)$$

and $S_k^{n+1} = S_k(t_{n+1})$. The additional volumetric flux constrains imposed on the instantaneous velocity field are

$$\Phi_k(\mathbf{u}^{n+1}) = Q_k(t_{n+1}) \equiv Q_k^{n+1} \quad , \quad k = K_P+1,..,K \quad (3.6)$$

We will show that the solution of the flow problem (3.3)-(3.6) can be obtained in the form of linear superposition of solutions to appropriately defined auxiliary Stokes problems. These problems, however, should be defined differently for proposed variants of the inlet/outlet conditions. We will begin with the solution method for the variant A.

To this aim, consider first the pair $\mathbf{w}_j \in V, \zeta_j \in Q$ which solves the following variational Stokes problem

$$\begin{cases} \dfrac{\alpha_0}{\Delta t}\int_\Omega \mathbf{w}_j \cdot \mathbf{v}\, d\Omega + \nu \int_\Omega \nabla \mathbf{w}_j : \nabla \mathbf{v}\, d\Omega - \int_\Omega \zeta_j \nabla \cdot \mathbf{v}\, d\Omega + \Phi_j(\mathbf{v}) = 0 \quad, \quad \forall \mathbf{v} \in V \\ \int_\Omega q \nabla \cdot \mathbf{w}_j\, d\Omega = 0 \quad, \quad \forall q \in L^2(\Omega) \end{cases} \qquad (3.7)$$

Define also the quantities $F_{jk} = \Phi_k(\mathbf{w}_j)$. Clearly, for each pair of indices $(j,k)$, $j,k = 1,..,K$ the number $F_{jk}$ is equal to the flow rate of the vector field $\mathbf{w}_j$ through the inlet/outlet section $\Gamma_k$. Since the flow in incompressible, for each $j = 1,..,K$ the equality $\sum_{k=1}^{K} F_{jk} = 0$ must hold.

In addition to the auxiliary problems (3.7), one needs to solve the following variational Stokes problem

$$\begin{cases} \dfrac{\alpha_0}{\Delta t}\int_\Omega \mathbf{w}_0 \cdot \mathbf{v}\, d\Omega + \nu \int_\Omega \nabla \mathbf{w}_0 : \nabla \mathbf{v}\, d\Omega - \int_\Omega \zeta_0 \nabla \cdot \mathbf{v}\, d\Omega = \mathbf{G}^{n+1} \quad, \quad \forall \mathbf{v} \in V \\ \int_\Omega q \nabla \cdot \mathbf{w}_0\, d\Omega = 0 \quad, \quad \forall q \in L^2(\Omega) \end{cases} \qquad (3.8)$$

where the right-hand side vector $\mathbf{G}^{n+1}$ is defined by the variant A of the formula (3.5). We also define the quantities $F_{0k} = \Phi_k(\mathbf{w}_0)$, $k = 1,..,K$, i.e., the volumetric flow rates of the vector field $\mathbf{w}_0$ through the inlet/outlet sections $\Gamma_k, k = 1,..,K$. Again, due to incompressibility constrain, $\sum_{k=1}^{K} F_{0k} = 0$.

The solution to the constrained Stokes problem (3.3)-(3.6) is sought in the form of the linear combinations of the auxiliary Stokes problems (3.7) and (3.8), namely

$$\mathbf{u}^{n+1} = \mathbf{w}_0 + \sum_{j=1}^{K} \beta_j \mathbf{w}_j \quad , \quad p^{n+1} = \zeta_0 + \sum_{j=1}^{K} \beta_j \zeta_j \tag{3.9}$$

After insertion of (3.9) to the first variation equality in (3.3) one obtains

$$\frac{\alpha_0}{\Delta t}\int_\Omega \mathbf{w}_0 \cdot \mathbf{v}\,d\Omega + \nu\int_\Omega \nabla\mathbf{w}_0 : \nabla\mathbf{v}\,d\Omega - \int_\Omega \zeta_0 \nabla\cdot\mathbf{v}\,d\Omega +$$
$$+ \sum_{k=1}^{K} \beta_k [\frac{\alpha_0}{\Delta t}\int_\Omega \mathbf{w}_j \cdot \mathbf{v}\,d\Omega + \nu\int_\Omega \nabla\mathbf{w}_j : \nabla\mathbf{v}\,d\Omega - \int_\Omega \zeta_j \nabla\cdot\mathbf{v}\,d\Omega ] + \tag{3.10}$$
$$+ \sum_{k=1}^{K} \eta_k [\Phi_k(\mathbf{w}_0) + \sum_{j=1}^{K} \beta_j \Phi_k(\mathbf{w}_j)]\Phi_k(\mathbf{v}) + \sum_{k=1}^{K} S_k \Phi_k(\mathbf{v}) = \mathbf{G}^{n+1}$$

where $\eta_k = \lambda_k + \frac{\alpha_0}{\Delta t}\gamma_k$. With the use of (3.7) and (3.8), the equality (3.10) reduces to

$$\sum_{k=1}^{K}[-\beta_k + \eta_k \Phi_k(\mathbf{w}_0) + \sum_{j=1}^{K}\beta_j \eta_k \Phi_k(\mathbf{w}_j) + S_k]\Phi_k(\mathbf{v}) = 0 \quad , \quad \forall \mathbf{v} \in V \tag{3.11}$$

The equality (3.11) should hold for each test function $\mathbf{v} \in V$, hence the following algebraic equations must be satisfied

$$-\beta_k + \sum_{j=1}^{K}\eta_k F_{jk}\beta_j + S_k + F_{0k}\gamma_k = 0 \quad , \quad k = 1,..,K \tag{3.12}$$

Additionally, the volumetric flow rate conditions (2.3) imply that

$$\sum_{j=1}^{K} F_{jk}\beta_j + F_{0k} = Q_k \quad , \quad k = K_P + 1,..,K \tag{3.13}$$

With the use of (3.13), last $K - K_P$ equations of the system (3.12) can be re-written as

$$-\beta_k + \eta_k Q_k + S_k = 0 \quad , \quad k = K_P + 1,..,K$$

Hence

$$S_k = \beta_k - \eta_k Q_k, \quad k = K_P + 1, .., K \tag{3.14}$$

and the final form of the linear algebraic system for unknown coefficients $\{\beta_j, j = 1, .., K\}$ reads

$$\begin{cases} \beta_k - \sum_{j=1}^{K} \eta_k F_{jk} \beta_j = S_k + F_{0k} \eta_k, & k = 1, .., K_P \\ \sum_{j=1}^{K} F_{jk} \beta_j = Q_k - F_{0k}, & k = K_P + 1, .., K \end{cases} \tag{3.15}$$

After the system (3.15) is solved, the unknown quantities $S_k, k = K_P + 1, .., K$, follow from the formulae (3.14).

Note that if all coefficients $\lambda_k, \gamma_k, k = 1, .., K$ are zero then the dissipative inlet/outlet conditions reduce to the "standard" open (aka pseudo-traction) conditions

$$\begin{cases} \beta_k = S_k, & k = 1, .., K_P \\ \sum_{j=K_P+1}^{K} F_{jk} \beta_j = Q_k - F_{0k} - \sum_{j=1}^{K_P} F_{jk} S_j, & k = K_P + 1, .., K \\ S_k = \beta_k, & k = K_P + 1, .., K \end{cases} \tag{3.16}$$

Note also that the relations (3.14) hold true also for $k = 1, .., K_P$, hence the inlet/outlet mean pressures at the time instant $t = t_{n+1}$ are

$$\bar{p}_k^{n+1} = S_k + \eta_k \Phi_k(\mathbf{u}^{n+1}) = \beta_k, \quad k = 1, .., K \tag{3.17}$$

which explains the physical meaning of the coefficients $\beta_k, j = 1, .., K$.

Consider now the solution method for the constrained Stokes problem with the inlet/outlet conditions in the variant B, i.e., defined by the formula (2.2b). The corresponding variational form of the semi-discretized Stokes problem (3.3) can be re-written as

$$\begin{cases} \dfrac{\alpha_0}{\Delta t}\int_\Omega \mathbf{u}^{n+1}\cdot\mathbf{v}\,d\Omega + \sum_{k=1}^{K}\hat{\eta}_k \int_{\Gamma_k}(\mathbf{u}^{n+1}\cdot\mathbf{n})(\mathbf{v}\cdot\mathbf{n})\,dA + \nu\int_\Omega \nabla\mathbf{u}^{n+1}:\nabla\mathbf{v}\,d\Omega - \int_\Omega p^{n+1}\nabla\cdot\mathbf{v}\,d\Omega + \\ +\sum_{k=1}^{K} S_k^{n+1}\Phi_k(\mathbf{v}) = \mathbf{G}^{n+1} \quad , \quad \forall \mathbf{v}\in V \\ \int_\Omega q\nabla\cdot\mathbf{u}\,d\Omega = 0 \quad , \quad \forall q \in L^2(\Omega) \end{cases} \qquad (3.18)$$

where $\hat{\eta}_k = \hat{\lambda}_k + \dfrac{\alpha_0}{\Delta t}\hat{\gamma}_k$ and the right-hand side vector $\mathbf{G}^{n+1}$ is defined by the variant B of the formula (3.5).

Again, the solution to the problem (3.18) with the flux rate constrains (2.3) can be constructed via superposition of particular solutions, accordingly to the formula (3.9). However, the auxiliary Stokes problems must be defined differently. In particular, the pairs $\mathbf{w}_j \in V, \zeta_j \in L^2(\Omega)$, $j=1,..K$, should now solve the following variational problems:

$$\begin{cases} \dfrac{\alpha_0}{\Delta t}\int_\Omega \mathbf{w}_j\cdot\mathbf{v}\,d\Omega + \sum_{k=1}^{K}\hat{\eta}_k\int_{\Gamma_k}(\mathbf{w}_j\cdot\mathbf{n})(\mathbf{v}\cdot\mathbf{n})\,dA + \nu\int_\Omega \nabla\mathbf{w}_j:\nabla\mathbf{v}\,d\Omega - \int_\Omega \zeta_j\nabla\cdot\mathbf{v}\,d\Omega + \\ +\Phi_j(\mathbf{v}) = 0 \quad , \quad \forall \mathbf{v}\in V \\ \int_\Omega q\nabla\cdot\mathbf{w}_j\,d\Omega = 0 \quad , \quad \forall q \in L^2(\Omega) \end{cases} \qquad (3.19)$$

The pair $\mathbf{w}_0 \in V, \zeta_0 \in L^2(\Omega)$ needs to be the solution to the following Stokes problem:

$$\begin{cases} \dfrac{\alpha_0}{\Delta t}\int_\Omega \mathbf{w}_0\cdot\mathbf{v}\,d\Omega + \sum_{k=1}^{K}\hat{\eta}_k\int_{\Gamma_k}(\mathbf{w}_0\cdot\mathbf{n})(\mathbf{v}\cdot\mathbf{n})\,dA + \nu\int_\Omega \nabla\mathbf{w}_0:\nabla\mathbf{v}\,d\Omega - \\ -\int_\Omega \zeta_0\nabla\cdot\mathbf{v}\,d\Omega = \mathbf{G}^{n+1} \quad , \quad \forall \mathbf{v}\in V \\ \int_\Omega q\nabla\cdot\mathbf{w}_0\,d\Omega = 0 \quad , \quad \forall q \in L^2(\Omega) \end{cases} \qquad (3.20)$$

The coefficients $\beta_k, k=1,..,K$ in the linear combinations (3.9) are determined from the same linear system (3.15), where all quantities $\eta_k$ should be replaced by the quantities $\hat{\eta}_k$. Then, the "far-field" pressures $S_k, k=K_P+1,..,K$, follow from the formulae (3.14). Also, upon the above replacement, the formulae (3.17) for the mean pressures at all inlets/outlets holds true. Note also that, in contrast to the variant A, the mass matrix obtained from a finite or spectral elements method applied to the variant B will include contributions from the boundary terms. Moreover, these additional terms cause coupling between all Cartesian velocity components and thus destroy a block-diagonal (in case of spectral element methods – even purely diagonal) structure of the mass matrix. Hence, numerical solution of the flow problem with inlet/outlet conditions in the variant B will be more complex than in the variant A.

## 4. Generalization to the Navier-Stokes flow

The numerical approach developed in Sections 2 and 3 can be applied to the Navier-Stokes system providing that a boundary-value problem to be solved in each time step is linear. Clearly, this condition is fulfilled when the nonlinear (convective) terms are treated in a fully explicit manner. Two most popular approaches are:
- approximation of the convective terms by a linear extrapolation formula of sufficiently high order,
- application of the Operator-Integration-Factor Splitting (OIFS) technique.

In the remaining part of this paper, we will focus on the latter technique as it usually guaranties better stability properties than the linear extrapolation. We will restrict our considerations to the flow problem formulation with the inlet/outlet conditions defined by the formula (2.2a), i.,e., to the variant A. The variant B can be treated analogously.

Consider the flow in the domain $\Omega$ governed by the Navier-Stokes and continuity equations

$$\begin{cases} \partial_t \mathbf{u} + (\mathbf{u} \cdot \nabla)\mathbf{u} - \nu \Delta \mathbf{u} + \nabla p = \mathbf{f} \\ \nabla \cdot \mathbf{u} = 0 \end{cases} \tag{4.1}$$

with the boundary and inlet/outlet conditions

$$\begin{cases} \mathbf{u}|_{\Gamma_0} = \mathbf{0} \\ p\mathbf{n} - \nu \frac{\partial}{\partial \mathbf{n}} \mathbf{u} - \left[\lambda_k \Phi_k(\mathbf{u}) + \gamma_k \partial_t \Phi_k(\mathbf{u})\right] \mathbf{n} = S_k \mathbf{n} \quad at \ \Gamma_k, \ k=1,..,K \end{cases} \quad (4.2)$$

where, as before, $S_k$, $k=1,..,K_P$, are given functions of time. The functions $S_k$, $k=K_P+1,..,K$ are to be found as a part of the solution, so that the velocity field $\mathbf{u}$ fulfills the flow rate conditions

$$\Phi_k(\mathbf{u}) = \int_{\Gamma_k} \mathbf{u} \cdot \mathbf{n} \, dA = Q_k(t) \quad , \quad k = K_P+1,..,K \quad (4.3)$$

The weak form of the flow problem (4.1)-(4.2) can be written as follows

(4.4)
$$\begin{cases} \int_\Omega \partial_t \mathbf{u} \cdot \mathbf{v} \, d\Omega + \sum_{k=1}^{K} \gamma_k \Phi_k(\partial_t \mathbf{u}) \Phi_k(\mathbf{v}) + \int_\Omega \mathbf{v} \cdot (\nabla \cdot \mathbf{u})\mathbf{u} \, d\Omega + \nu \int_\Omega \nabla \mathbf{u} : \nabla \mathbf{v} \, d\Omega - \int_\Omega p \nabla \cdot \mathbf{v} \, d\Omega + \\ + \sum_{k=1}^{K} \left[\lambda_k \Phi_k(\mathbf{u}) + S_k\right] \Phi_k(\mathbf{v}) = \int_\Omega \mathbf{f} \cdot \mathbf{v} \, d\Omega \quad , \quad \forall \mathbf{v} \in V \\ \int_\Omega q \nabla \cdot \mathbf{u} \, d\Omega = 0 \quad , \quad \forall q \in Q \end{cases}$$

The computational procedure for the time-step update of the solution to above problem consists of three steps:

**Step 1: The OIFS**

For each $j=1,..,J$, one integrates over the time interval $[t_{n+1-j}, t_{n+1}]$ the following (weakly-posed) initial-value problem (IVP), using the explicit 4$^{th}$-order Runge-Kutta method:

$$\begin{cases} \frac{d}{dt}\left[\int_\Omega \hat{\mathbf{u}}_j \cdot \mathbf{v} \, d\Omega + \sum_{k=1}^{K} \gamma_k \Phi_k(\hat{\mathbf{u}}_j) \Phi_k(\mathbf{v})\right] = -\int_\Omega \mathbf{v} \cdot (\hat{\mathbf{u}}_j \cdot \nabla)\hat{\mathbf{u}}_j \, d\Omega \quad , \quad \forall \mathbf{v} \in V \\ \hat{\mathbf{u}}_j \big|_{t=t_{n+1-j}} = \mathbf{u}^{n+1-j} \end{cases} \quad (4.5)$$

The rationale behind the IVP (4.5) is provided in the Appendix. One should notice the additional terms in the left side of the (4.5), which appear due to inertial parts of the inlet/outlet conditions and lead to coupling of all Cartesian components of the field $\hat{\mathbf{u}}_j$. As a result of this coupling, the mass matrix obtained after discretization of the system (4.5) with finite elements is not block-diagonal anymore. If the spectral element are applied, the effect of coupling between all Cartesian directions due to inertial nature of the inlet/outlet conditions is particularly severe – the mass matrix loses its usual purely diagonal structure. As a result, the ODE system obtained after spatial discretization of (4.5) is not explicit with respect to time derivatives of the unknowns. This fact renders the OIFS step computationally more complex in comparison to ordinary dissipative (meaning, with $\gamma_k = 0, k = 1,..,K$) or open inlet/outlet conditions.

The computed terminal values $\hat{\mathbf{u}}_l^{n+1} \equiv \hat{\mathbf{u}}_l(t_{n+1})$, $j = 1,..,J$, should be stored to be used in the next step of the solution update procedure. Typically, the RK4 integration step is $\Delta t_{RK4} = \Delta t / m_{RK4}$, where $\Delta t$ denotes the main time step of the flow simulation and $m_{RK4}$ is a small integer number (say, between 2 and 5).

**Step 2: Solution of the Stokes problem**

In the second sub-step, one determines the pair $(\mathbf{u}^{n+1}, p^{n+1}) \in V \times L^2(\Omega)$ and the quantities $S_k^{n+1}$, $k = K_P + 1,..,K$, which satisfy the following variational Stokes problem

$$\begin{cases} \dfrac{\alpha_0}{\Delta t}\int_\Omega \mathbf{u}^{n+1} \cdot \mathbf{v}\,d\Omega + \sum_{k=1}^{K}\eta_k \Phi_k(\mathbf{u}^{n+1})\Phi_k(\mathbf{v}) + \nu\int_\Omega \nabla \mathbf{u}^{n+1} : \nabla \mathbf{v}\,d\Omega - \int_\Omega p^{n+1}\nabla \cdot \mathbf{v}\,d\Omega + \\ + \sum_{k=1}^{K} S_k^{n+1}\Phi_k(\mathbf{v}) = \mathbf{G}_{OIFS}^{n+1} \quad , \quad \forall \mathbf{v} \in V \\ \int_\Omega q\nabla \cdot \mathbf{u}^{n+1}\,d\Omega = 0 \quad , \quad \forall q \in Q \end{cases} \quad (4.6)$$

and the flowrate constrains

$$\Phi_k(\mathbf{u}^{n+1}) := \int_{\Gamma_k} \mathbf{u}^{n+1} \cdot \mathbf{n}\,dS = Q_k(t_{n+1}) \equiv Q_k^{n+1} \quad , \quad k = K_P + 1,..,K. \quad (4.7)$$

The right-hand side vector in (4.6) is defined by the formula

$$\mathbf{G}_{OIFS}^{n+1} = \int_{\Omega} \mathbf{f}^{n+1} \cdot \mathbf{v} \, d\Omega + \frac{1}{\Delta t} \sum_{j=1}^{J} \alpha_j \int_{\Omega} \hat{\mathbf{u}}_j^{n+1} \cdot \mathbf{v} \, d\Omega + \frac{1}{\Delta t} \sum_{j=1}^{J} \alpha_j \sum_{k=1}^{K} \gamma_k \Phi_k(\hat{\mathbf{u}}_j^{n+1}) \Phi_k(\mathbf{v}) \quad (4.8)$$

Note that the only difference between the Stokes problems (4.6)-(4.8) and (3.3)-(3.6) is the slight modification in the right-hand side functions. Indeed, the past values of the velocity field $\mathbf{u}^{n+1-j}$, $j=1,..,J$, which appear in the second term of the formula (3.5) (in both variants) are replaced in the formula (4.9) by the vector fields $\hat{\mathbf{u}}_j^{n+1}$ computed in the OIFS sub-step. Still, the full solution $(\mathbf{u}^{n+1}, p^{n+1})$, satisfying the flowrate constrains (4.3), can be obtained as a linear combination of the auxiliary Stokes flows $(\mathbf{w}_j, \zeta_j)$, $j=1,..,K$, obtained from (3.18) and the Stokes flow $(\mathbf{w}_0, \zeta_0)$ obtained from (3.20) with the function $\mathbf{G}^{n+1}$ replaced by $\mathbf{G}_{OIFS}^{n+1}$.

## 5. Further generalization to nonlinear inertial-dissipative conditions

In this section, we consider the inertial-dissipative inlet/outlet conditions augmented by the quadratic terms as follows

(5.1a)

(A) $p\mathbf{n} - \nu \frac{\partial}{\partial \mathbf{n}} \mathbf{u} - \mathbf{n}[(\lambda_k + \gamma_k \frac{d}{dt}) \Phi_k(\mathbf{u}) + \chi_k |\Phi_k(\mathbf{u})| \Phi_k(\mathbf{u})] = S_k \mathbf{n}$ at $\Gamma_k$, $k=1,..,K$

(5.1b)

(B) $p\mathbf{n} - \nu \frac{\partial}{\partial \mathbf{n}} \mathbf{u} - \mathbf{n}[(\hat{\lambda}_k + \hat{\gamma}_k \partial_t)\mathbf{u} \cdot \mathbf{n} + \hat{\chi}_k |\mathbf{u} \cdot \mathbf{n}|(\mathbf{u} \cdot \mathbf{n})] = S_k \mathbf{n}$ at $\Gamma_k$, $k=1,..,K$

This extension renders both variants of the conditions (5.1) nonlinear. Let us again take a closer look at the variant A (5.1a). Following the arguments presented in Section 2.2 we obtain the formula with a section-averaged pressure $\bar{p}_k$, namely

$$\bar{p}_k - S_k = \gamma_k \frac{d}{dt} \Phi_k(\mathbf{u}) + \lambda_k \Phi_k(\mathbf{u}) + \chi_k |\Phi_k(\mathbf{u})| \Phi_k(\mathbf{u}) \quad (5.2)$$

The formula (5.2) can be interpreted as an unsteady Darcy-Forchheimer law for a hypothetical porous structures from which fluid is sucked through the *k-th* inlet to the flow domain or to which the fluid spills out through the *k-th* outlet. This flow setting is schematically depicted in Figure 2. Using nonlinearly augmented inlet/outlet conditions may be regarded as more adequate if a fluid motion is sufficiently rapid and for this reason inertia-based resistance effects should be also accounted for.

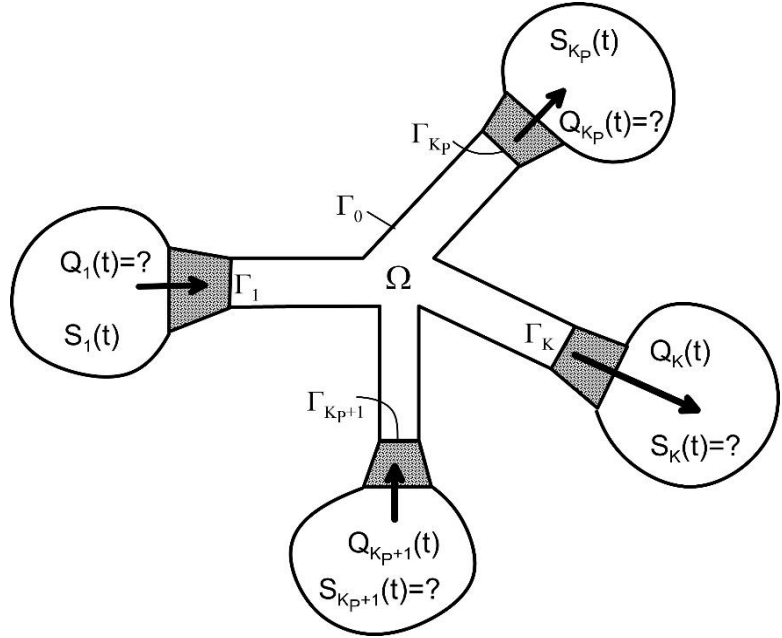

*Figure 2. Conceptual flow domain with inlet/outlet linked through porous interfaces with external containers.*

As before, the initial/boundary value problem with the boundary conditions (5.1) should be posed in an appropriate variational form. For the variant A this form reads:

*Find:* $\mathbf{u} \in V$, $p \in L^2(\Omega)$ *and functions* $\{S_{K_P+1}(t),..,S_K(t)\}$ *such that the equalities*

$$\begin{cases} \int_\Omega \partial_t \mathbf{u} \cdot \boldsymbol{v}\, d\Omega + \int_\Omega \boldsymbol{v} \cdot (\mathbf{u} \cdot \nabla)\mathbf{u}\, d\Omega + \nu \int_\Omega \nabla \mathbf{u} : \nabla \boldsymbol{v}\, d\Omega - \int_\Omega p \nabla \cdot \boldsymbol{v}\, d\Omega + \\ + \sum_{k=1}^{K} [\gamma_k \frac{d}{dt}\Phi_k(\mathbf{u}) + \lambda_k \Phi_k(\mathbf{u}) + \chi_k |\Phi_k(\mathbf{u})|\underline{\Phi_k(\mathbf{u})} + S_k]\Phi_k(\boldsymbol{v}) = \int_\Omega \mathbf{f} \cdot \boldsymbol{v}\, d\Omega \\ \int_\Omega q \nabla \cdot \mathbf{u}\, d\Omega = 0 \end{cases} \quad (5.3)$$

*hold true for each* $\boldsymbol{v} \in V$ *and* $q \in L^2(\Omega)$.

Again, it also assumed that $\Phi_k(\mathbf{u}) = Q_k(t)$ for $k = K_P + 1,..,K$, and the functions $\{S_1(t),..,S_{K_P}(t)\}$ are given. Note that the nonlinear terms in (5.3) are underlined.

The solution method proposed for the problem (5.3) is a straightforward extension of the technique described in Sections 3 and 4. The additional nonlinear terms can be conveniently included in the OIFS sub-step. To this aim, the right-hand sides of the auxiliary initial-value problems (4.6) are extended to contains also the nonlinear boundary term

$$\begin{cases} \frac{d}{dt}[\int_\Omega \hat{\mathbf{u}}_j \cdot \mathbf{v}\,d\Omega + \sum_{k=1}^{K} \gamma_k \Phi_k(\hat{\mathbf{u}}_j)\Phi_k(\mathbf{v})] = \\ -\int_\Omega \mathbf{v}\cdot(\nabla\cdot\hat{\mathbf{u}}_j)\hat{\mathbf{u}}_j\,d\Omega - \sum_{k=1}^{K}\chi_k |\Phi_k(\hat{\mathbf{u}}_j)|\Phi_k(\hat{\mathbf{u}}_j)\Phi_k(\mathbf{v}) \quad,\quad \forall \mathbf{v}\in V \\ \hat{\mathbf{u}}_j\big|_{t=t_{n+1-j}} = \mathbf{u}^{n+1-j} \end{cases} \quad (5.4)$$

As explained in Section 4, the vector fields $\hat{\mathbf{u}}_l^{n+1} \equiv \hat{\mathbf{u}}_l(t_{n+1})$, $j=1,..,J$, are computed by means of the explicit 4$^{th}$-order Runge-Kutta scheme and then used in the variational Stokes problem (4.7) solved at each time step. The flow rate constrains (4.8) are resolved by superposing particular Stokes solutions defined in Section 3.

The solution procedure for the variant B of the conditions (5.1) can be constructed along the same lines. The OIFS sub-step assumes the following form

$$\begin{cases} \frac{d}{dt}\int_\Omega \hat{\mathbf{u}}_j \cdot \mathbf{v}\,d\Omega = -\int_\Omega \mathbf{v}\cdot(\hat{\mathbf{u}}_j\cdot\nabla)\hat{\mathbf{u}}_j\,d\Omega - \sum_{k=1}^{K}\hat{\chi}_k\int_{\Gamma_k}|\hat{\mathbf{u}}_j\cdot\mathbf{n}|(\hat{\mathbf{u}}_j\cdot\mathbf{n})(\mathbf{v}\cdot\mathbf{n})\,dA \quad,\quad \forall \mathbf{v}\in V \\ \hat{\mathbf{u}}_j\big|_{t=t_{n+1-j}} = \mathbf{u}^{n+1-j} \end{cases} \quad (5.5)$$

Again, the fields $\hat{\mathbf{u}}_l^{n+1} \equiv \hat{\mathbf{u}}_l(t_{n+1})$, $j=1,..,J$, are computed by means of the explicit 4$^{th}$-order Runge-Kutta scheme and then used in the variational Stokes problem

$$\begin{cases} \frac{\alpha_0}{\Delta t}\int_\Omega \mathbf{u}^{n+1}\cdot\mathbf{v}\,d\Omega + \sum_{k=1}^{K}\hat{\eta}_k\int_{\Gamma_k}(\mathbf{u}^{n+1}\cdot\mathbf{n})(\mathbf{v}\cdot\mathbf{n})\,dA + \nu\int_\Omega \nabla\mathbf{u}^{n+1}:\nabla\mathbf{v}\,d\Omega - \int_\Omega p^{n+1}\nabla\cdot\mathbf{v}\,d\Omega + \\ +\sum_{k=1}^{K}S_k^{n+1}\Phi_k(\mathbf{v}) = \mathbf{G}_{OIFS}^{n+1} \quad,\quad \forall \mathbf{v}\in V \\ \int_\Omega q\nabla\cdot\mathbf{u}^{n+1}\,d\Omega = 0 \quad,\quad \forall q\in Q \end{cases} \quad (5.6)$$

where $\hat{\eta}_k = \hat{\lambda}_k + \frac{\alpha_0}{\Delta t}\hat{\gamma}_k$ and

$$\mathbf{G}_{OIFS}^{n+1} = \int_{\Omega} \mathbf{f}^{n+1} \cdot \boldsymbol{v} \, d\Omega + \tfrac{1}{\Delta t} \sum_{j=1}^{J} \alpha_j \int_{\Omega} \hat{\mathbf{u}}_j^{n+1} \cdot \boldsymbol{v} \, d\Omega + \tfrac{1}{\Delta t} \sum_{j=1}^{J} \alpha_j \sum_{k=1}^{K} \hat{\gamma}_k \int_{\Gamma_k} (\hat{\mathbf{u}}_j^{n+1} \cdot \mathbf{n})(\boldsymbol{v} \cdot \mathbf{n}) \, dA \qquad (5.7)$$

The solution to (5.6) which satisfies the flow rate constrains (4.7) is found as a superposition of auxiliary Stokes flows, as described in Section 3.

**6. Posing the flow problem with nonlinear inertial-dissipative conditions as an optimal control problem**

In this Section, we derive an alternative formulation of the flow problem with nonlinear DF-like inlet/outlet conditions and flow rate constrains. We will pose this flow problem as an open-loop optimal control problem and use the continuous adjoint technique as a tool for determination of the generalized gradient of a goal functional with respect to control variables. The goal functional to be minimized within the control time interval $[0,T]$ is defined as follows

$$\mathcal{J} = \tfrac{1}{2} \int_0^T \sum_{k=K_P+1}^{K} [\int_{\Gamma_k} \boldsymbol{u} \cdot \boldsymbol{n} \, dS - Q_k(t)]^2 \, dt \qquad (6.1)$$

Clearly, this functional measures – in a least squares sense – deviation of actual volumetric fluxes through the inlet/outlet sections $\Gamma_k, k = K_P + 1,..,K$ from assumed flow rate variations $Q_k(t), k = K_P + 1,..,K$. This deviation is to be minimized by appropriate time variations of the "far-field" pressure $S_k(t), k = K_P + 1,..,K$ corresponding to these inlet/outlets. The functions $S_k(t), k = 1,..,K_P$ are assumed given and fixed.

The "state constrains" are imposed on the above control task, namely, the velocity field $\boldsymbol{u}$ must solve the Navier-Stokes system together with the initial and boundary conditions formulated in Section 5. In order to solve such constrained optimal control problem, we introduce the augmented Lagrangian as follows

$$\mathcal{H} = \underbrace{\frac{1}{2}\int_0^T \sum_{k=K_P+1}^{K} [\int_{\Gamma_k} \boldsymbol{u}\cdot\boldsymbol{n}\,dS - Q_k(t)]^2\,dt}_{\mathcal{J}} +$$
$$+ \underbrace{\int_0^T\int_\Omega [\partial_t \boldsymbol{u} + (\boldsymbol{u}\cdot\nabla)\boldsymbol{u} + \nabla p - \nu\Delta \boldsymbol{u}]\cdot \boldsymbol{w}\,d\Omega\,dt - \int_0^T\int_\Omega (\nabla\cdot\boldsymbol{u})q\,d\Omega\,dt}_{\mathcal{L}}$$
(6.2)

Note that if the pair $(\boldsymbol{u}, p)$ satisfies equations of the Navier-Stokes system then $\mathcal{H} = \mathcal{J}$. The vector field $\boldsymbol{w}$ is referred to as the adjoint velocity, while the scalar field $q$ is called the adjoint pressure. The further derivation steps include:

- calculation the formula for the first variation $\delta\mathcal{H}$ of the augmented Lagrangian,
- transformation of this formula to such form that the variations $\delta\boldsymbol{u}$ and $\delta p$ stand free of any differentia operator,
- defining the adjoint flow problem for the pair $(\boldsymbol{w}, q)$ such that all terms containing explicitly variations $\delta\boldsymbol{u}$ and $\delta p$ vanish.

The final outcome of the above procedure is the expression of $\delta\mathcal{H}$ which includes only variations of the control variables. The generalized gradients can be inferred directly from this expression and used any gradient-based optimization algorithm.

Clearly, $\delta\mathcal{H} = \delta\mathcal{J} + \delta\mathcal{L}$. The first variation of the goal functional $\mathcal{J}$ is equal

$$\delta\mathcal{J} = \int_0^T \left[ \sum_{k=K_P+1}^{K} \int_{\Gamma_k} E_k(t)\boldsymbol{n}\cdot\delta\boldsymbol{u}\,dS \right] dt \qquad (6.3)$$

where

$$E_k(t) = \int_{\Gamma_k} \boldsymbol{u}\cdot\boldsymbol{n}\,dS - Q_k(t), \quad k = K_P + 1,..,K \qquad (6.4)$$

The calculation of $\delta\mathcal{L}$ is more complicated. In the first step, one obtains

$$\delta \mathcal{L} = \int_0^T \int_\Omega [\partial_t \delta u + (\delta u \cdot \nabla) u + (u \cdot \nabla) \delta u + \nabla \delta p - \nu \Delta \delta u] \cdot w \, d\Omega \, dt -$$
$$- \int_0^T \int_\Omega (\nabla \cdot \delta u) q \, d\Omega \, dt$$
(6.5)

Next, the variations $\delta u$ and $\delta p$ should be freed from action of differential operators. To this aim, integration by parts, the Gauss-Green-Ostrogradsky theorem and the 2nd Green identity are used. Since the initial condition for the velocity field are fixed then $\delta u|_{t=0} = 0$. Moreover, the velocity vanishes identically at the solid part of the boundary and thus $\delta u|_\Gamma = 0$. Eventually, the formula for $\delta \mathcal{L}$ reads

(6.6)

$$\delta \mathcal{L} = \int_\Omega w(T) \cdot \delta u(T) \, dx + \int_0^T \int_\Omega [-\partial_t w - (u \cdot \nabla) w + (\nabla u)^T w - \nu \Delta w + \nabla q] \cdot \delta u \, dx \, dt +$$
$$+ \int_0^T \int_\Omega (-\nabla \cdot w) \delta p \, dx \, dt + \int_0^T \sum_{k=1}^K \int_{\Gamma_k} [\nu \tfrac{\partial}{\partial n} w - qn + (u \cdot n) w] \cdot \delta u \, dS \, dt +$$
$$+ \int_0^T \sum_{k=1}^K \int_{\Gamma_k} w \cdot [\delta p n - \nu \tfrac{\partial}{\partial n} \delta u] \, dS \, dt$$

Special attention should be paid to the last term in (6.6). Let us choose the inlet/outlet conditions in the variant A, i.e., defined by Eq. (5.1a). Hence, we have at $\Gamma_k$, $k = 1,..,K$

$$p\mathbf{n} - \nu \tfrac{\partial}{\partial \mathbf{n}} \mathbf{u} = [S_k(t) + (\lambda_k + \gamma_k \tfrac{d}{dt}) \Phi_k(u) + \chi_k |\Phi_k(\mathbf{u})| \Phi_k(\mathbf{u})] \mathbf{n}$$
(6.7)

Let us recall that the functions $S_k(t), k = 1,..K_P$ are given while the functions $S_k(t), k = K_P + 1,..,K$ serve as the control variables. Thus, the first variation of the conditions (6.7) can be expressed as follows

(6.8)

$$\delta p\mathbf{n} - \nu \tfrac{\partial}{\partial \mathbf{n}} \delta u =$$
$$= \begin{cases} [(\lambda_k + \gamma_k \tfrac{d}{dt}) \Phi_k(\delta u) + 2\chi_k |\Phi_k(u)| \Phi_k(\delta u)] \mathbf{n} \, , \, k = 1,..,K_P \\ [\delta S_k(t) + (\lambda_k + \gamma_k \tfrac{d}{dt}) \Phi_k(\delta u) + 2\chi_k |\Phi_k(u)| \Phi_k(\delta u)] \mathbf{n} \, , \, k = K_P + 1,..,K \end{cases}$$

After insertion of the expressions (6.8), integration by parts and using $\delta u|_{t=0} = \mathbf{0}$, one obtains the following form of the last term in the formula (6.6)

$$\int_0^T \int_{\partial\Omega} \mathbf{w} \cdot [\delta p \mathbf{n} - \nu \tfrac{\partial}{\partial \mathbf{n}} \delta \mathbf{u}] dS\, dt =$$
$$= \int_0^T \left\{ \sum_{k=1}^{K} [(\lambda_k - \gamma_k \tfrac{d}{dt}) \Phi_k(\mathbf{w}) + 2\chi_k |\Phi_k(\mathbf{u})| \Phi_k(\mathbf{w})] \Phi_k(\delta \mathbf{u}) \right\} dt + \qquad (6.9)$$
$$+ \sum_{k=1}^{K} \gamma_k \Phi_k[\mathbf{w}(T)] \Phi_k[\delta \mathbf{u}(T)] + \int_0^T \left[ \sum_{k=K_P+1}^{K} \Phi_k(\mathbf{w}) \delta S_k(t) \right] dt$$

Finally, the variation $\delta \mathcal{H}$ takes the following form

(6.10)

$$\delta H = \int_0^T \int_\Omega [-\partial_t \mathbf{w} - (\mathbf{u} \cdot \nabla)\mathbf{w} + (\nabla \mathbf{u})^T \mathbf{w} - \nu \Delta \mathbf{w} + \nabla s] \cdot \delta \mathbf{u}\, d\mathbf{x}\, dt +$$
$$+ \int_0^T \int_\Omega (-\nabla \cdot \mathbf{w}) \delta p\, d\mathbf{x}\, dt + \int_\Omega \mathbf{w}(T) \cdot \delta \mathbf{u}(T)\, d\mathbf{x} + \sum_{k=1}^{K} \gamma_k \Phi_k[\mathbf{w}(T)] \Phi_k[\delta \mathbf{u}(T)] +$$
$$+ \int_0^T \left[ \sum_{k=1}^{K_P} \int_{\Gamma_k} \mathbf{b}_k \cdot \delta \mathbf{u}\, dS \right] dt + \int_0^T \left[ \sum_{k=K_P+1}^{K} \int_{\Gamma_k} \{\mathbf{b}_k + E_k(t)]\mathbf{n}\} \cdot \delta \mathbf{u}\, dS \right] dt +$$
$$+ \int_0^T \left[ \sum_{k=K_P+1}^{K} \Phi_k(\mathbf{w}) \delta S_k(t) \right] dt$$

where, for $k = 1,.., K$

(6.11)

$$\mathbf{b}_k = [\nu \tfrac{\partial}{\partial \mathbf{n}} \mathbf{w} - q\mathbf{n} + (\mathbf{u} \cdot \mathbf{n})\mathbf{w}]\Big|_{\Gamma_k} + [(\lambda_k - \gamma_k \tfrac{d}{dt}) \Phi_k(\mathbf{w}) + 2\chi_k |\Phi_k(\mathbf{u})| \Phi_k(\mathbf{w})] \mathbf{n}$$

In the next step, the adjoint problem is defined for the pair $(\mathbf{w}, q)$. This problem should be formulated in such a way that the only term remaining in the formula (6.10) is the last term containing variations of the control variables. Other terms should be eliminated since they contain dependent variations of the solution components $\mathbf{u}$ and $p$. Thus, the adjoint velocity $\mathbf{w}$ and the adjoint pressure $q$ must satisfy the linear differential system

$$\begin{cases} -\partial_t \boldsymbol{w} - (\boldsymbol{u} \cdot \nabla)\boldsymbol{w} + (\nabla \boldsymbol{u})^T \boldsymbol{w} = -\nabla q + \nu \Delta \boldsymbol{w} \\ \nabla \cdot \boldsymbol{w} = 0 \end{cases} \quad \text{in } \Omega \qquad (6.12)$$

as well as the boundary conditions

$$\begin{cases} \boldsymbol{w}|_\Gamma = \boldsymbol{0} \\ \boldsymbol{b}_k = 0 \text{ at } \Gamma_k, k = 1,..,K_P \\ \boldsymbol{b}_k = -E_k(t)\boldsymbol{n} \text{ at } \Gamma_k, k = K_P + 1,..,K \end{cases} \qquad (6.13)$$

where $E_k(t)$, $k = K_P + 1,..,K$, are defined by (6.4). The above system needs to be integrated backward in time, with the terminal condition $\boldsymbol{w}(T) = \boldsymbol{0}$.

Once the solution to the adjoint problem (6.12), (6.13) is found, the formula (6.10) for the variation $\delta \mathcal{H}$ reduces to

$$\delta H = \int_0^T \left[ \sum_{k=K_P+1}^{K} \Phi_k(\boldsymbol{w}) \delta S_k(t) \right] dt \qquad (6.14)$$

From above one can infer that the generalized gradients of the augmented Lagrangian with respect to the control variables are equal to the volumetric flow rates of the adjoint velocity through the inlets/outlets $\Gamma_k, k = K_P + 1,..,K$

$$\frac{\partial \mathcal{H}}{\partial S_k} = \Phi_k(\boldsymbol{w}) \equiv \int_{\Gamma_k} \boldsymbol{w} \cdot \boldsymbol{n}\, dS \ , \ k = K_P + 1,..,K \qquad (6.15)$$

If the steepest descent method is used, the updates of the control variables $S_k(t), k = K_P + 1,..,K$ in the $(n+1)-th$ iteration are computed as follows

$$S_k^{(n+1)} = S_k^{(n)} - \varepsilon^{(n)} \frac{\partial \mathcal{H}^{(n)}}{\partial S_k} = S_k^{(n)} - \varepsilon^{(n)} \int_{\Gamma_k} \boldsymbol{w}^{(n)} \cdot \boldsymbol{n}\, dS \qquad (6.16)$$

where $\varepsilon^{(n)}$ is a small positive number.

In order to apply the finite or spectral elements, the adjoint system (6.12), (6.13) should be posed in a weak work. The variational formulation of this problem reads:

Find $w \in V = \{v : v \in H_1(\Omega) \wedge v|_\Gamma = \mathbf{0}\}$ and $s \in L^2(\Omega)$ such that the equalities

$$\begin{cases} -\int_\Omega \partial_t w \cdot \xi \, d\Omega - \int_\Omega \xi \cdot (u \cdot \nabla) w \, d\Omega + \int_\Omega \xi \cdot (\nabla u)^T w \, d\Omega + \nu \int_\Omega \nabla w : \nabla \xi \, d\Omega \\ -\int_\Omega q \nabla \cdot \xi \, d\Omega = -\sum_{k=1}^K [(\lambda_k - \gamma_k \frac{d}{dt}) \Phi_k(w) + 2\chi_k |\Phi_k(u)| \Phi_k(w)] \int_{\Gamma_k} \xi \cdot n \, dS \\ -\sum_{k=1}^K \int_{\Gamma_k} (u \cdot n) w \cdot \xi \, dS - \sum_{k=K_P+1}^K E_k(t) \int_{\Gamma_k} \xi \cdot n \, dS \\ (\nabla \cdot w, \zeta) = 0 \end{cases} \quad (6.17)$$

are satisfied for each $\xi \in V$ and each $\zeta \in L^2(\Omega)$.

A few remarks should be made regarding the variational problem (6.17) and its numerical solution:

1. An alternative form of the adjoint Navier-Stokes equation can be obtained if the following transformation of the term $\int_\Omega w \cdot (\delta u \cdot \nabla) u \, d\Omega$ is applied

$$\int_\Omega w \cdot (\delta u \cdot \nabla) u \, d\Omega = \int_\Omega \frac{\partial u_i}{\partial x_j} w_i \, \delta u_j \, d\Omega = \int_\Omega \frac{\partial}{\partial x_j} (u_i w_i \, \delta u_j) \, d\Omega - \int_\Omega u_i \frac{\partial}{\partial x_j} (w_i \, \delta u_j) \, d\Omega =$$

$$= \int_{\partial\Omega} u_i w_i n_j \delta u_j \, dS - \int_\Omega \frac{\partial w_i}{\partial x_j} u_i \, \delta u_j \, d\Omega - \int_\Omega u_i w_i \delta(\frac{\partial}{\partial x_j} u_j) \, d\Omega =$$

$$= \int_{\partial\Omega} (u \cdot w) n \cdot \delta u \, dS - \int_\Omega \delta u \cdot (\nabla w)^T u \, d\Omega$$

After insertion to the formula for the first variation of the augmented Lagrangian the following form of the adjoint problem is obtained

$$\begin{cases} -\partial_t w - [\nabla w + (\nabla w)^T] u = -\nabla q + \nu \Delta w \\ \nabla \cdot w = 0 \end{cases} \quad in \; \Omega \quad (6.18)$$

$$\begin{cases} \boldsymbol{w}|_{\Gamma} = \boldsymbol{0} \\ \hat{\boldsymbol{b}}_k = 0 \ at\ \Gamma_k, k=1,..,K_P \\ \hat{\boldsymbol{b}}_k = -E_k(t)\boldsymbol{n}\ at\ \Gamma_k, k=K_P+1,..,K \end{cases} \quad (6.19)$$

where $\hat{\boldsymbol{b}}_k = \boldsymbol{b}_k + (\boldsymbol{u}\cdot\boldsymbol{w})\boldsymbol{n}|_{\Gamma_k}$, $k=1,..,K$ and $\boldsymbol{b}_k$ are defined by (6.11). The corresponding variational formulation reads:

*Find* $\boldsymbol{w} \in V = \{\boldsymbol{v}: \boldsymbol{v} \in H_1(\Omega) \wedge \boldsymbol{v}|_{\Gamma} = \boldsymbol{0}\}$ *and* $s \in L^2(\Omega)$ *such that the equalities*

$$\begin{cases} -\int_{\Omega} \partial_t \boldsymbol{w} \cdot \boldsymbol{\xi}\, d\Omega - \int_{\Omega} \boldsymbol{\xi}\cdot(\boldsymbol{u}\cdot\nabla)\boldsymbol{w}\, d\Omega - \int_{\Omega} \boldsymbol{\xi}\cdot(\nabla\boldsymbol{w})^T \boldsymbol{u}\, d\Omega + \nu \int_{\Omega} \nabla\boldsymbol{w}:\nabla\boldsymbol{\xi}\, d\Omega \\ -\int_{\Omega} q\nabla\cdot\boldsymbol{\xi}\, d\Omega = -\sum_{k=1}^{K}[(\lambda_k - \gamma_k \tfrac{d}{dt})\Phi_k(\boldsymbol{w}) + 2\chi_k|\Phi_k(\boldsymbol{u})|\Phi_k(\boldsymbol{w})]\int_{\Gamma_k} \boldsymbol{\xi}\cdot\boldsymbol{n}\, dS \\ -\sum_{k=1}^{K}\int_{\Gamma_k}[(\boldsymbol{u}\cdot\boldsymbol{n})\boldsymbol{w} + (\boldsymbol{u}\cdot\boldsymbol{w})\boldsymbol{n}]\cdot\boldsymbol{\xi}\, dS - \sum_{k=K_P+1}^{K} E_k(t)\int_{\Gamma_k} \boldsymbol{\xi}\cdot\boldsymbol{n}\, dS \\ (\nabla\cdot\boldsymbol{w},\zeta)=0 \end{cases} \quad (6.20)$$

*are satisfied for each* $\boldsymbol{\xi} \in V$ *and each* $\zeta \in L^2(\Omega)$.

The formulas (6.14) and (6.15) remain unchanged.

2. No matter which variational form is used, the adjoint problem is obviously linear. Note also that the adjoint convective terms always lead to coupling of all three Cartesian directions. Additional source of this coupling are, as in the primal problem, the boundary terms. In general, this issue would require some attention in terms of an efficient numerical implementation, especially in 3D cases. For instance, one might consider a possibility of using OIFS methodology, i.e., an approach normally applied to nonlinear terms in the primal problem. Unconditional stability will be most certainly lost in such semi-implicit approach. Besides, the unsteady boundary terms still spoil the otherwise nice structure of the mass matrix, as pointed out in Chapter 4. More detailed discussion of the implementation- related issued will be presented in the second part of this paper.

3. Note that the primal velocity field $\boldsymbol{u}$ appears in the adjoint variational equalities (6.10) and (6.20). Thus, in order to compute the adjoint solution, the primal velocity field must be available over the whole control period $[0,T]$. Since storing full history of the primal solution may require, especially in 3D cases, a prohibitive amount of computer memory, an appropriate checkpointing strategy is compulsory. Moreover, in the context of realistic 3D application, an appropriate checkpointing technique should be suitable for parallel computations and dynamic time step adaptation. Again, this issue will be pursued in more details in the second part of this work.

4. It is possible to pose the optimal control problem with the other variant of nonlinear inertial-dissipative conditions, i.e., the one defined by (5.1b). Derivations of the corresponding adjoint problem - very similar to those above – are left to the Reader.

## 7. Summary

In this paper, we have formulated the extension of the dissipative inlet/outlet conditions, proposed originally as a part of flow models for the respiratory system. The extension consists in adding the inertia term, which seems physically justified, especially for rapidly pulsating or oscillatory flows. Two variants of inertia-augmented dissipative inlet/outlet conditions have been considered. The numerical method for the Stokes problem with both variants of the novel inlet/outlet conditions and additional flowrate constrains has been also constructed. The general idea of this method is to take advantage of linearity and use superposition of appropriately defined auxiliary Stokes problems in order to build the full solution at a given time instant. It has been shown that the variational form of the boundary value problems for the auxiliary solutions depends on the variant of inlet/outlet conditions. It has been also shown that the proposed numerical method can be easily generalized for the Navier-Stokes system. To this aim, the Operator-Integration-Factor-Splitting technique has been applied to the nonlinear (convective) terms, which effectively reduces the procedure of the flow update to the linear Stokes problem. Hence, only slight modifications in the right-hand side vectors are needed to obtain the solution to the nonlinear flow problem.

It has been already pointed out that generalization of the proposed superposition-based numerical method to Navier-Stokes flows is conditioned by a treatment of the convective terms. In the current work, these terms are integrated separately by fully explicit RK4 method. For the large-scale 3D applications, the memory requirements of the OIFS sub-step based on the "naive" version of this method may be prohibitive. Therefore, for demanding applications more sophisticated, low-storage variants of RK methods are recommended (see for instance [22,23]).

More radical alternative is switching to semi-implicit methods, which typically enjoy better stability properties and admit larger time steps. Specifically, the convective term can be approximated by the formula $(\mathbf{u}^{n+1} \cdot \nabla)\mathbf{u}^{n+1} \approx (\mathbf{u}_{ext} \cdot \nabla)\mathbf{u}^{n+1}$, where $\mathbf{u}_{ext}$ denotes the vector field obtained via linear extrapolation from the flow history, for instance $\mathbf{u}_{ext} = 2\mathbf{u}^n - \mathbf{u}^{n-1}$. If such approximation is applied to the Navier-Stokes problem (4.1)-(4.3), the solution via superposition of special solutions in a time step will be still possible. There are, however, two main issues which make the semi-implicit method not as much appealing. First, determination of each auxiliary pair $(\mathbf{w}_j, \zeta_j)$, $j = 0,..,J$, will require solution of the Oseen boundary-value problem, which are more difficult task than solving the Stokes problems. Moreover, all pairs $(\mathbf{w}_j, \zeta_j)$ will depend on the flow history, meaning that $J+1$ Oseen problems need to be solved in each time step of the flow simulation. Note that in the methods proposed in the Sections 3 and 4 only one Stokes problem for the pair $(\mathbf{w}_0, \zeta_0)$ is solved in each time step, while the pairs $(\mathbf{w}_j, \zeta_j)$, $j = 1,..,J$, can be determined prior to the main simulation.

A radically different approach to the consider class of problems is to re-formulate them as the open-loop optimal control problems. This approach has been presented in the Section 6, where we have formulated the optimal control problem for the Navier-Stokes system supplemented with the nonlinear inertial dissipative conditions of the Darcy-Forchheimer type. Next, we have derived the adjoint problem (both in strong and variational forms) and explained how to calculate the generalized gradient of the goal functional with respect to the control variables, i.e., the time-dependent "far-field" pressures linked to selected inlets/outlets of the flow domain. Finally, we have made a few general comments of the numerical implementation of the proposed methodology.

**Appendix: Operator-Integration-Factor-Splitting method**

Here, we present briefly the derivation of the OIFS technique for a dynamical system with a nontrivial "mass" operator. Consider the evolutionary equation in the operator form

$$M \frac{d}{dt} \mathbf{v} = L\mathbf{v} + N(\mathbf{v}) \tag{A1}$$

where $M$ and $L$ are linear operators and $N(\mathbf{v})$ is the nonlinear term. Consider the time-dependent operator $F(t_*;t)$ (parametrized by $t_*$) such that

$$\frac{d}{dt}[F(t_*;t)M\mathbf{v}(t)] = F(t_*;t)L\mathbf{v}(t) \tag{A2}$$

Assume also that $F(t_*;t_*) = id$, i.e., this operator reduces to identity when $t = t_*$. It follows from the Eq. (A1) and the postulated property (A2) that

$$[\frac{d}{dt}F(t_*;t)]M\mathbf{v}(t) + F(t_*;t)N[\mathbf{v}(t)] = 0 \tag{A3}$$

Consider the following initial value problem (again, parametrized with $t_*$)

$$\begin{cases} \frac{d}{d\tau}M\hat{\mathbf{v}}(t_*;\tau) = N[\hat{\mathbf{v}}(t_*;\tau)] \\ \hat{\mathbf{v}}(t_*;\tau = t) = \mathbf{v}(t) \end{cases} \tag{A4}$$

Then, from (A3) and (A4), one obtains

$$\begin{aligned}\frac{d}{d\tau}[F(t_*;\tau)M\hat{\mathbf{v}}(t_*;\tau)] &= [\frac{d}{d\tau}F(t_*;\tau)]M\hat{\mathbf{v}}(t_*;\tau) + F(t_*;\tau)M\frac{d}{d\tau}\hat{\mathbf{v}}(t_*;\tau) = \\ &= -F(t_*;\tau)N[\hat{\mathbf{v}}(t_*;\tau)] + F(t_*;\tau)N[\hat{\mathbf{v}}(t_*;\tau)] = 0\end{aligned} \tag{A5}$$

Hence $F(t_*;\tau)M\hat{\mathbf{v}}(t_*;\tau) = const$ and

$$F(t_*;t)M\hat{\mathbf{v}}(t_*;t) = F(t_*;t_*)M\hat{\mathbf{v}}(t_*;t_*) = M\hat{\mathbf{v}}(t_*;t_*) \tag{A6}$$

The formula (A6) has the following meaning: in order to calculate the action of the operator $F(t_*;t)M$ on the function $\mathbf{v}(t) = \hat{\mathbf{v}}(t_*,t)$ one needs to determine the solution to the initial value problem (A4) at the time $\tau = t_*$.

In order to construct a time integration scheme for the equation (A1) one can apply the $J^{th}$-order backward differentiation formula (BDF) to the equation (A2) with $t_* = t_{n+1}$. This formula can be written as

$$\frac{1}{\Delta t}[\alpha_0 F(t_{n+1};t_{n+1})M\mathbf{v}_{n+1} - \sum_{j=1}^{J}\alpha_k F(t_{n+1};t_{n+1-j})M\mathbf{v}_{n+1-j}] = L\mathbf{v}_{n+1} \tag{A7}$$

where $\alpha_0, \alpha_1, ..., \alpha_J$ are the known coefficients. In order to avoid explicit reference to the operator $F(t_*;t_{n+1-j})$, $j = 1,..,J$, the following initial value problems must be solved

$$\begin{cases} \dfrac{d}{dt} M\hat{\mathbf{v}}^j(t) = N[\hat{\mathbf{v}}^j(t)] \;,\; t \in [t_{n+1-j}, t_{n+1}] \\ \hat{\mathbf{v}}^j(t = t_{n+1-j}) = \mathbf{v}(t_{n+1-j}) \end{cases} ,\quad j = 1,..,J \tag{A8}$$

Using (A6) one obtains the values $M\hat{\mathbf{v}}_{n+1}^k \equiv M\hat{\mathbf{v}}^k(t_{n+1}) = F(t_{n+1}; t_{n+1-k}) M\mathbf{v}(t_{n+1-j})$, $j = 1,..,J$. Using also the identity $F(t_{n+1}; t_{n+1}) = id$, the formula (A7) can be re-written as

$$\frac{\alpha_0}{\Delta t} M\mathbf{v}_{n+1} - L\mathbf{v}_{n+1} = \frac{1}{\Delta t} \sum_{j=1}^{J} \alpha_j M\hat{\mathbf{v}}_{n+1}^j \tag{A9}$$

In the CFD context, the solution of the initial value problems (B8) and (B9) involves numerical integration of the convective terms, typically by means of the $4^{th}$-order explicit Runge-Kutta method which combines reasonable numerical cost with good stability properties in convective-dominated problems. Low-storage variants of this method [22,23] may be used for large problems to reduce the memory requirements.